\documentstyle[11pt,revpasp,epsf]{article}
\def\edcomment#1{\iffalse\marginpar{\raggedright\sl#1\/}\else\relax\fi}
\reversemarginpar

\begin{document}
\title{Abundance Ratios in Early-Type Galaxies}
\author{Reynier Peletier}
\affil{Dept. of Physics, University of Durham, South Road, Durham, DH1 3LE, UK}

\begin{abstract}
Although evidence is accumulating that abundance ratios in galaxies are
often non-solar, they are far from understood. I resume the current
evidence for non-solar abundance ratios, supplementing the recent review
by Worthey (1998) with some new results. It appears that the Mg/Fe abundance ratio only
depends on the mass of the galaxy, not on the formation time-scale.
For massive galaxies [Mg/Fe] $>$ 0, while small galaxies show solar abundance
ratios. 
Information about abundances of other element is scarce, but new
evidence is given that [Ca/Fe] is solar, or slightly lower than solar,
contrary to what is expected for an $\alpha$-element.
\end{abstract}

\section{Introduction}

The distribution of element abundances in galaxies is an important fossil 
record of their formation and evolution. Of primary importance is 
the metallicity  (the mass fraction of all
elements heavier than He), which contains important information about
the past star formation history, which could well be strongly influenced
by mergers and interactions. For early-type galaxies, the metallicity is
an excellent estimator of the luminosity, or even the mass. 
Also very important are metallicity 
gradients, which are among the few parameters that can tell us something about
the orbital structure in galaxies. 

In recent years the quality of observational data has become so good,
that one can also start thinking of measuring abundances of individual
elements in external galaxies. Individual abundances will greatly 
help to understand their chemical evolution. In particular, one will 
be able to understand better
the way in which the ISM of galaxies is enriched by metals, and 
what the relevant time-scales are. In the last 2 decades it has become
clear that the abundance distribution in stars is not always the same
as in the Sun. This was discovered first in individual stars in our
galaxy (see Wheeler, Sneden \& Truran 1989) and later in integrated
spectra of elliptical galaxies (Peletier 1989, Worthey, Faber \& Gonz\'alez 1992). 
In this review I will discuss what we currently know about abundance
ratios in galaxies, and what we can learn from this about the formation
and evolution of galaxies. This paper is in part based on the excellent paper
by Worthey (1998), but also includes some new high-quality data, which might
shed new light on some issues in this rapidly evolving field.

The paper starts in Section 2 discussing some global relations for galaxies 
as a function of luminosity or velocity dispersion. In Section 3 the 
central abundances and abundance ratios are discussed, and in Section 4
their gradients. In Section 5 it is briefly discussed how we can 
understand non-solar abundance ratios.  The need for
stellar models with non-solar abundance ratios is mentioned in Section 6,
after which some conclusions are given.

\section{Global Relations}

It has been known for some time that the stellar populations of 
early-type galaxies are strongly linked to their other properties.
Two examples are the colour-magnitude relation (see e.g. Sandage \&
Visvanathan 1978) and the relation between Mg$_2$ and velocity dispersion
($\sigma$)
(Terlevich et al. 1981). Both relations can be understood well if
the average metallicity of a galaxy is larger when the galaxy is brighter.
The usefulness of these relations in understanding galaxies 
strongly increased when Schweizer
\& Seitzer (1992) showed that the residuals of the colour/Mg$_2$ -- $\sigma$
relation correlate with a parameter indicating recent mergers. This
correlation implies that the scatter  for undisturbed 
early-type galaxies is smaller than the amount due to observational 
uncertainties, while
colours of disturbed galaxies are somewhat bluer for a given $\sigma$,
due to increased star formation during the merger.
This interpretation was confirmed by the work of Bower, Lucey \& Ellis (1992),
who found a very low scatter in the colour-magnitude relation (CMR) in the
Coma cluster, showing no sign of any recent star formation in the early-type
galaxies in this cluster. Later, using the more sensitive H$\gamma$ absorption
line index, Caldwell et al. (1993) showed that many early-type galaxies
in the SW part of the cluster show signs of small amounts of young stars.

Recently, A. Terlevich (1998) redid the study of Bower et al. (1992) with a 
significantly larger sample in the Coma cluster. He finds that the intrinsic 
scatter in the elliptical galaxies in $U-V$ is 0.036 mag (consistent with Bower
et al.). No difference was found in the slope of the CMR between ellipticals
and S0 galaxies. The slope of the CMR was also the same in different
areas of the cluster. Most galaxies blue-ward of the CMR are either 
late-type galaxies (Andreon et al. 1996) or seem to have late-type
morphologies. In the outer parts of the cluster the residuals about the
CMR become somewhat bluer, in agreement with the the results of Caldwell 
et al. (1993).
These detailed colour studies indicate that the colour of an early-type
galaxy is determined by its luminosity to a high accuracy. The same can be said
of the Mg$_2$ absorption line (Bender, Burstein \& Faber 1993, Schweizer 
\& Seitzer 1992).  
Although a colour of a galaxy, and also an absorption line, 
depends on many parameters, like metallicity, age and Initial Mass
Function (IMF) slope, the fact that the CMR has the same slope 
across various factors of 10 in luminosity implies that the relations
described above are almost certainly driven by metallicity: fainter 
galaxies have lower metallicities, and because of that bluer colours.
This behaviour has been successfully reproduces in 
Galactic enrichment models (e.g. Arimoto 
\& Yoshii 1987). Starting with a proto-galactic cloud, several generations
of stars are formed, until the rate of Supernovae is so large that all the gas
is expulsed from the galaxy, and the star formation stops.  
Being able to model the CMR has been very important for our understanding
of galaxy formation. We are now however in a position that we can go one step
further, and ask ourselves how the abundances of individual elements varies 
as a function of $\sigma$ or luminosity.

We know that the strength of the Mg$_2$ feature
depends strongly on the Mg abundance (Worthey 1998), so the 
fact that there is a good correlation between Mg$_2$ and $\sigma$ or luminosity
tells us that the Mg-abundance 
increases with that parameter. Colours don't contain much information
about individual element-abundances, except that blue colours, 
through line blanketing, depend very much on the abundance of Fe-peak
elements, which would imply that the Fe-elements would be
a strong function as well of $\sigma$. Do we know whether 
the abundances of other elements
also correlate strongly with $\sigma$?

High-quality measurements of elements other than Mg are scarce, because 
of the high signal-to-noise required, and the difficulty associated with 
calibrating indices like $\langle$Fe$\rangle$ and H$\beta$ onto the Lick system
(Faber et al. 1985, Gorgas et al. 1993, Worthey et al. 1994). 
Recently Fisher, Franx \& Illingworth (1996) and J\o rgensen (1997) published 
$\langle$Fe$\rangle$ and $\sigma$ data for a reasonably large number 
of galaxies. Although in both cases the scatter in the $\langle$Fe$\rangle$
-- $\sigma$ relation is large, it is smaller in the data of J\o rgensen (1997).
In Fig.~1a we show her Figure 2, displaying Mg$_2$, $\langle$Fe$\rangle$ 
and H$\beta$ as a function
of $\sigma$. If, as she claims, the scatter is larger than the instrumental
scatter, it would mean that $\langle$Fe$\rangle$ cannot be a simple
function of metallicity, like Mg$_2$, but that there has to be a second
parameter, most
likely the Mg/Fe abundance ratio. This parameter cannot be the age, 
because the galaxies in J\o rgensen (1997) are early-type galaxies in clusters,
which fall well onto their CMR. Her interpretation however might not be correct.
Recently, Kuntschner (1998) published some high-quality data of the
Fornax cluster (see also Kuntschner \& Davies 1998). As a comparison we
have plotted them in Fig.~1b. He showed that 
for $\log~\sigma_0$ $\ge$ 1.9 the 
scatter in Fe3', an index very similar to $\langle$Fe$\rangle$, and in
H$\beta$ is smaller or comparable to the scatter in Mg$_2$. This is 
a very important result, implying that in Fornax $\langle$Fe$\rangle$,
just like Mg$_2$, depends only on $\sigma$, not on a second parameter.
Since Kuntschner's signal-to-noise is much larger than J\o rgensen's,
it looks as if the observational scatter in J\o rgensen (1997) has been
underestimated. If this is not the case, it would mean that in some galaxy
clusters the stellar populations are affected by a second parameter,
while in others (Fornax) this would not be the case.

\begin{figure}[h]
\begin{center}
\mbox{\epsfxsize=13cm  \epsfbox{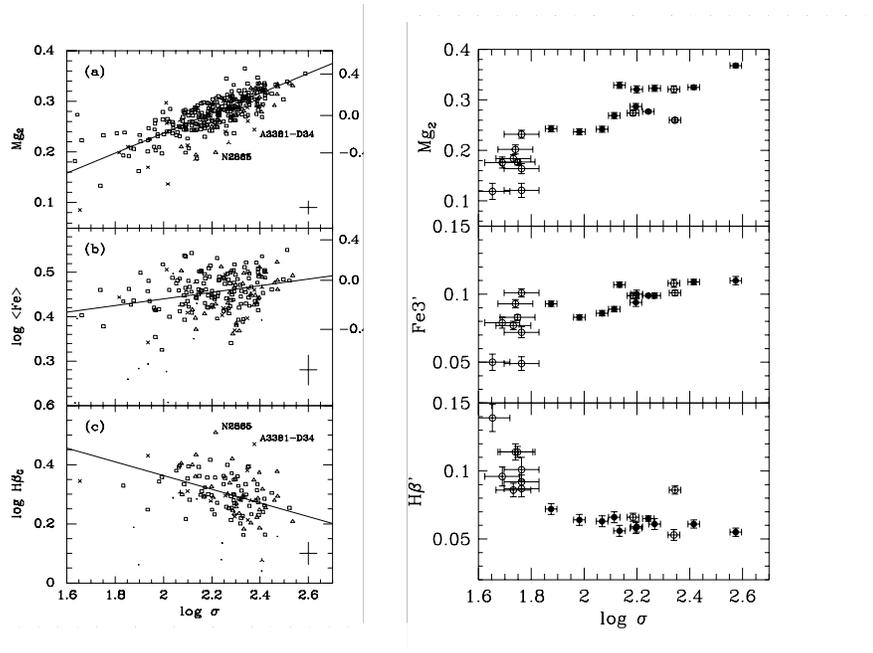}}
\caption{Line strength vs. $\sigma$ relations from J\o rgensen
(1997, left) for 9 clusters and Kuntschner (1998, right) for the Fornax cluster. 
In the figure on the right open symbols denote S0 galaxies, while 
ellipticals are indicated by filled circles. Fe3' is a pseudo-magnitude index
containing the Fe features at 4383, 5270 and 5335 \AA\ , and H$\beta$'
is the H$\beta$ index in magnitudes.}
\end{center}
\end{figure}

\section{Abundance Ratios in the Centres of Galaxies}

\subsection{Mg and Fe}

The largest dataset of nuclear line strengths in ellipticals and bulges
has been the Lick sample, which was finally published by Trager et al.
in 1998. It consists of measurements of almost 400 galaxies. The observations
were performed with the Lick IDS, the same instrument with which the
stars defining the Lick system have been observed. Worthey (1998) shows 
$\langle$Fe$\rangle$ vs. Mg$_2$ for a subset of this sample in his Figure 1.
In this Figure no difference can be seen between bulges of S0s and ellipticals.
Generally, objects with central velocity dispersion below about 200 km/s
seem to have solar Mg/Fe ratios, while the others are over-abundant in Mg.
Since the Lick IDS detector suffered from several instrumental problems,
the errors in the individual measurements are larger than one would like.
For that reason I have made a different compilation, with data from more
recent papers with high quality line strength measurements, and shown that
in Fig.~2. It includes a sample of spiral bulges (Jablonka, Martin \& 
Arimoto 1996) subdivided in a group of early-type bulges 
(type 0-2) and later type objects (type 3-5), a sample of bulges of S0 galaxies
(Fisher et al. 1996), a mixed sample of ellipticals
and S0 galaxies (Kuntschner 1998), a sample of mainly faint elliptical
galaxies (Halliday 1998) and two samples of bright ellipticals.
The models plotted here are from Vazdekis et al. (1996). The conclusion from 
Fig.~2 is again that bulges and ellipticals are indistinguishable. The
models of Vazdekis et al. (1996) are compatible with those of Worthey 
(1994), and also for these models galaxies start deviating from their 
locus at Mg$_2$ $>$ 0.25. Bulges of Sa-Sc spirals, which were not 
included in the Lick sample, seem to have solar Mg/Fe ratios. Although
there might be some small systematic offsets between the individual samples,
they generally agree well with each other.. There might be a few objects, which do not
follow the general trend. NGC~4458 and NGC~4464 (large filled dots, 
from Halliday 1998) have a very low Mg/Fe ratio, compared to other galaxies
with the same Mg$_2$. Both galaxies are faint ellipticals; NGC~4458
rotates very slowly, and has a small kinematically decoupled core
(Simien \& Prugniel 1998), while NGC~4464 
has a v/$\sigma$ of about 0.5
for an ellipticity of 0.3, so (v/$\sigma$)$^*$ is close to 1 (Davies 
et al. 1983), so that it is probably an oblate, rotating 
elliptical.

\begin{figure}[h]
\begin{center}
\mbox{\epsfxsize=10cm  \epsfbox{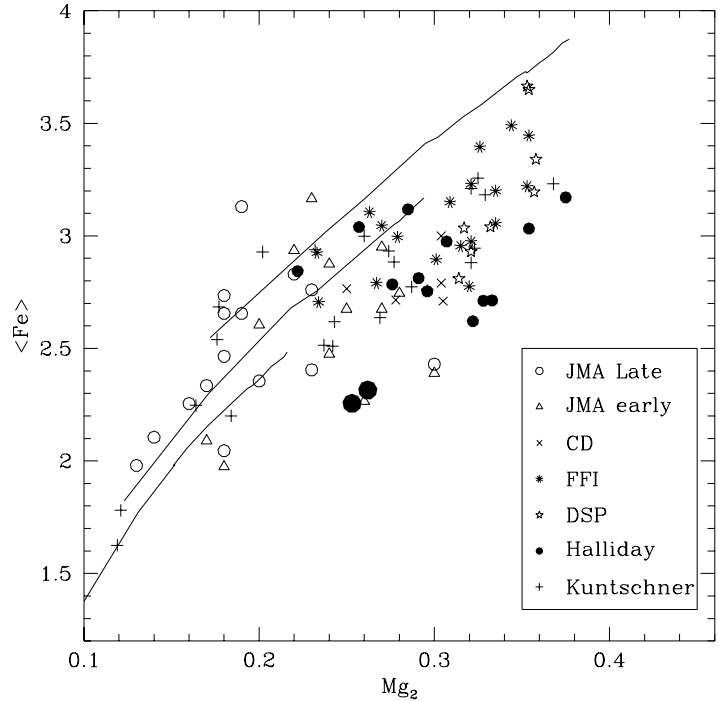}}
\caption{Literature compilation of Mg$_2$ vs. $\langle$Fe$\rangle$. Plotted
are central values. Also added are models with solar abundance ratios
of Vazdekis et al. (1996). The lines represent models of constant
metallicity (resp. Z=0.008, 0.02 and 0.05 from left to right), along which
age increases to the right. JMA = Jablonka et al. (19960; CD = Carollo,
Danziger \& Buson (1993); FFI = Fisher et al. (1996), and DSP = Davies et al.
(1993).}
\end{center}
\end{figure}

The situation for our own Galactic Bulge is in agreement with external
galaxies. It is estimated from high-resolution spectra that the mean
[Fe/H] in Baade's window is somewhat lower than solar (--0.25, McWilliam 
\& Rich 1994), but that [Mg/Fe] there is 0.3 -- 0.4.
For such a Mg-abundance, one would expect the Mg$_2$ index to be about 
0.29, assuming an age of 17 Gyr, if one uses the Vazdekis et al. (1996) 
models, or less, if the bulge would be younger. Assuming that the 
bulge is indeed old, it will lie in Fig.~2 in the region where galaxies
are over-abundant in Mg w.r.t. Fe. 

\subsection{Other elements}

Worthey (1998) extensively summarises our knowledge about the relative
abundance of metals other than Mg in giant ellipticals and in the 
Galactic Bulge. Although it appears that the situation concerning
the elements Sc, V and Ti is very complicated and confusing, there
are about half a dozen other elements for which we know something about
their behaviour in giant elliptical galaxies. In Table~1 I have 
schematically summarised our knowledge about them. All the information
has been obtained from various Lick indices, by comparing measured
line strengths with the values that one expects based on stellar population
models. As Worthey (1998) mentions, there are several differences in
element abundances
between giant ellipticals and our Bulge, which implies that there
must have been differences in their formation processes.
Especially in the case of N this is striking: while in the Bulge the
abundance of CN is generally lower than solar, in giant ellipticals
this is the opposite. It seems that [C/Fe] $\approx$ 0 in ellipticals 
(from the C$_2$4668 line) and in our Bulge (Worthey 1998), so that 
it is thought that N is depleted in our Galaxy and overabundant in
giant ellipticals. Peculiar is also that [O/Fe] $\approx$ 0
(McWilliam \& Rich 1994), since one would expect that O, an $\alpha$-element,
would follow Mg. McWilliam \& Rich (1994) however warn us that the only
stars for which they could measure O are at the tip of the RGB,
where their abundances might not be representative of the Bulge because 
of metal-enrichment in the star itself. More measurements of O-abundances 
would be very welcome.

\begin{table}[h]
\begin{center}
\caption{Abundance ratios for some important elements in giant ellipticals,
the Bulge, and the galactic halo}
\begin{tabular}{l|ccc}
\multicolumn{4}{c}{~} \\
\tableline
~ & \multicolumn{3}{c}{[X/Fe] in} \\
Element X & Giant Ell. & Gal. Bulge & Gal. Halo \\
\tableline
C & Solar & Solar ? & Solar \\
N & $>$0  & $<$0  & $\le$0 \\
O & Unknown & Solar ? & $>$0 \\
Mg & $>$0 & $>$0 & $>$0 \\
Na & $>$0 & $>$0 & $<$0 \\
Ca & $\le$0 ? & Solar & $>$0 \\ 
\tableline
\end{tabular}
\end{center}
\end{table}

Here I would like to revisit our ideas about the [Ca/Fe] abundance ratio
in giant ellipticals. The Lick system has two indices which can be used
to measure the Ca abundances in galaxies: Ca 4227 and possibly Ca 4455. Both
are faint, narrow features that are difficult to measure in giant 
ellipticals, because of the large correction for velocity broadening that
one has to apply to measure them. Vazdekis et al. (1997), using high signal to
noise spectra of three giant early-type galaxies, found that the 
measured Ca 4227 in all of them was much lower than expected from their
and also Worthey (1994)'s stellar population model. Their Fig.~13 nicely
illustrates how enormous the velocity dispersion correction is here. An
independent confirmation of this result comes from observations of
the NIR Ca II triplet (CaT) of the same 3 galaxies. We observed them
using 2d-FIS, an Integral Field Spectrograph on the 4.2m WHT at La Palma,
using a fibre bundle to feed the light from the Cassegrain focus to the 
slit of the ISIS double spectrograph (Peletier et al. 1999). Details 
about the instrument are
given in Arribas, Mediavilla, \& Rasilla (1991). In Fig.~3 our central 
measurements are plotted. Using the system
of D\'\i az, Terlevich \& Terlevich (1989) to define the band-passes, it was found that the 
CaT equivalent width is less strong than predicted by the models
of Vazdekis et al. (1996) with solar Ca/Fe ratios (see also Fig.~3). 
Although those models were based on the
stellar library of D\'\i az et al. (1989), which doesn't fully cover 
the range of metallicities of giant ellipticals, the main conclusions will
not change. Also, the same result is obtained if the models of
Garc\'\i a-Vargas, Moll\'a \& Bressan (1998) are used. Our observations of the CaT however 
are in good agreement with Terlevich, D\'\i az \& Terlevich (1990). 

How to explain this apparent {\sl under-abundance} of Ca? It is possibly
that the effect is entirely caused by a problem in calculating the models.
Idiart, Thevenin \& de Freitas Pacheco (1997) and also Borges et al. (1995) point out that most
models calculate integrated spectra by summing linear combinations
of observed spectra of standard stars, assuming that [Ca/Fe] = 0 for
those standard stars. Idiart et al. however 
obtained a separate library of standard stars,
determined Ca, Mg and Fe abundances for each star, and calculated 
integrated indices using those individual abundances. Applying their
models (Fig.~3) they find that [Ca/Fe] in the three galaxies is solar.
This is however still peculiar, since Ca is an $\alpha$-element, which
properties should follow closely those of Mg. Since the library of Idiart
et al. (1997) also is rather limited, it is very important to obtain
a stellar library of the size of the Lick library in the region of the
CaT, to be able to interpret this important line index, which can also
be used to constrain the IMF in galaxies. Together with the group at
the Universidad Complutense in Madrid we are currently working on providing
such a library (Cenarro et al. 1999, in preparation).

\begin{figure}[h]
\begin{center}
\mbox{\epsfxsize=13cm  \epsfbox{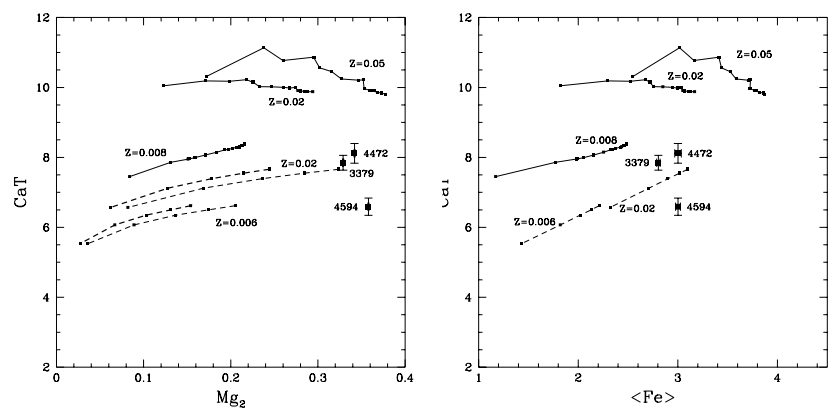}}
\caption{Models of the Ca T compared with the observations in the 
central aperture of 1.7$''$ (from Peletier et al. 1999). 
The drawn lines are models of Vazdekis et al. 
(1996), with metallicities of Z=0.008, 0.02 and 0.05, and with age ranging from
1 to 17 Gyr from left to right. The dashed lines are the models of
Idiart et al. (1997) and Borges et al. (1995). 
In the left figure the thick dashed lines indicate
models with solar [Mg/Fe], while for the thin dashed line [Mg/Fe]=0.30.}
\end{center}
\end{figure}

\section{Line Strength Gradients in Galaxies}

Line strength gradients have been presented by various authors 
(e.g. Gonz\'alez 1993, Davies, Sadler \& Peletier 1993, 
Carollo, Danziger \& Buson 1993,
Fisher, Franx \& Illingworth 1995, 1996, Vazdekis et al. 1997), but only for
the Mg$_2$ and Mg$_b$ indices reasonably high quality measurements are 
available in the literature for a sufficiently large sample, and possibly 
also for $\langle$Fe$\rangle$ and H$\beta$. Gradients in elliptical
galaxies in Fe and Mg indices are generally following tracks with 
constant [Mg/Fe], which means that for many galaxies the gradients
are steeper than the line linking galaxy nuclei (e.g. Worthey et al. 
1992, Davies et al. 1993). New, excellent quality data by Halliday
(1998, Figure~4) confirm this also for low luminosity ellipticals. Current data
seems to imply that Mg/Fe within all galaxies is constant. 
The behavior of Mg vs. Fe seems to be the same in bulges and disks.
Fisher et al. (1996) find that in S0 galaxies the radial gradients in 
Mg$_2$ and $\langle$Fe$\rangle$ are smaller along the major axis than
on the minor axis, implying that the gradients in the disk are smaller
than those in the bulge. However, in the Mg$_2$ vs. $\langle$Fe$\rangle$
diagram the galaxies have the same slope, on both major and minor axis.

\begin{figure}
\begin{center}
\mbox{\epsfxsize=13cm  \epsfbox{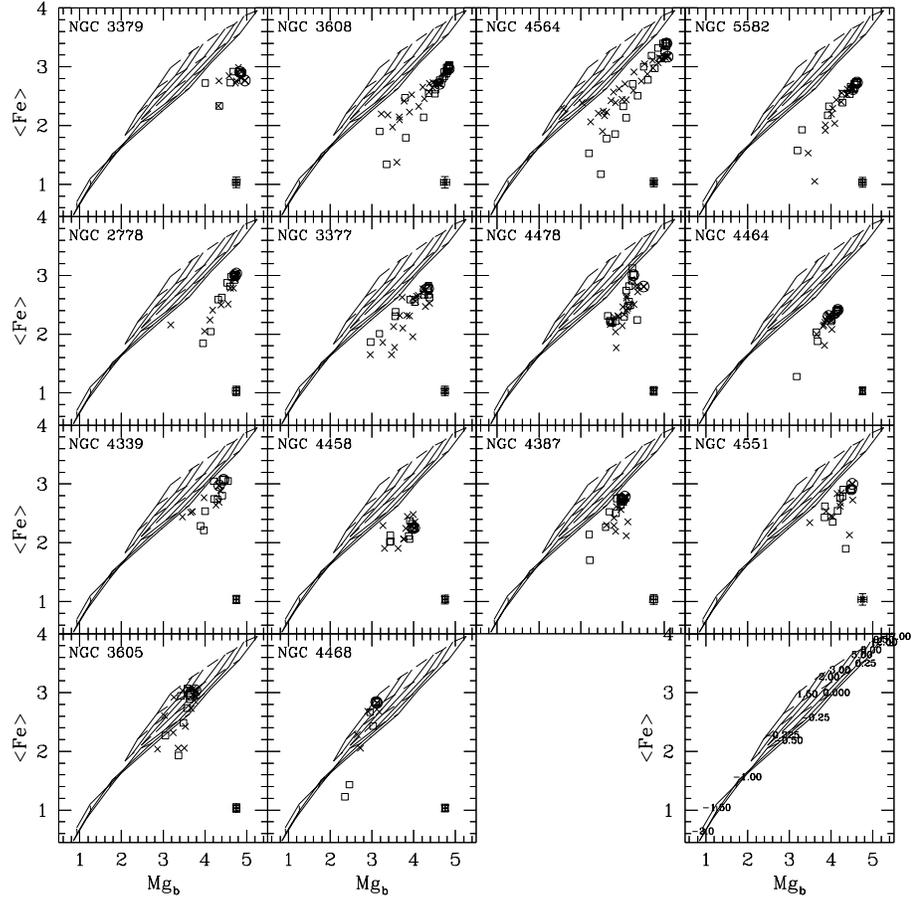}}
\caption{Measurements of Mg$_b$ and $\langle$Fe$\rangle$ for both the major
and minor axes of the sample of Halliday (1998) are compared with the 
models of Worthey (1994). Major axis measurements are given by crossed symbols
and minor axis by open squares. Central measurements for both axes are
encircled. Typical errors for Mg$_b$ and $\langle$Fe$\rangle$ are given in the
lower right-hand corner of each plot, the larger errors indicating
errors for the outer parts of each galaxy and smaller errors, for the
galaxy centre. The model grids consist of
dashed lines connecting points of equal metallicity and solid lines,
points of equal age; values of age are given in Gyr and metallicity,
in units of [Fe/H]. A guide plot is given in lower right
where lines corresponding to particular ages and metallicities are
labelled.}
\end{center}
\end{figure}

Information about abundance gradients in other elements is limited. Vazdekis
et al. (1997) present gradients in about 20 lines for the three above 
mentioned
galaxies. They find that the radial gradients can be explained well
by gradients in the overall metallicity. Much more work however has to
be done to investigate gradients in, e.g., bulges and smaller elliptical
galaxies.

Recently it has become possible to make reliable line strength maps
in two dimensions, using Integral Field Spectrography. This offers an 
exciting range of new possibilities. For example, in galaxies with
kinematically decoupled cores it can be investigated whether for example
Mg/Fe in the decoupled core is different from the ratio in the rest
of the galaxy, giving clues
to the origin of the decoupled core. In Peletier et al. (1999) one
galaxy with a decoupled core is included, the Sombrero galaxy, which has
a rapidly rotating inner disk (e.g. Wagner, Bender \& Dettmar 1988). In Fig.~5 we show
the Mg$_2$ and $\langle$Fe$\rangle$ maps in an inner field of 8.2$''$
$\times$ 11$''$. The continuum intensity is shown as well. The figure shows
that Mg is enhanced in the inner disk, compared to the bulge, and that
Fe is almost not enhanced (see also Emsellem et al. 1996). 
The noise however in the $\langle$Fe$\rangle$
image is still so large that it cannot be established whether the Mg/Fe
ratio itself in the inner disk has gone up. This galaxy shows, together
with other cases in the literature (e.g. NGC~4365, Surma \& Bender 1995,
NGC~7626, Davies et al. 1993) that the central Mg abundance in galaxies
with kinematically decoupled cores increases sharply, but up to now
Mg/Fe seems to remain constant. 

\begin{figure}[h]
\begin{center}
\mbox{\epsfysize=8cm  \epsfbox{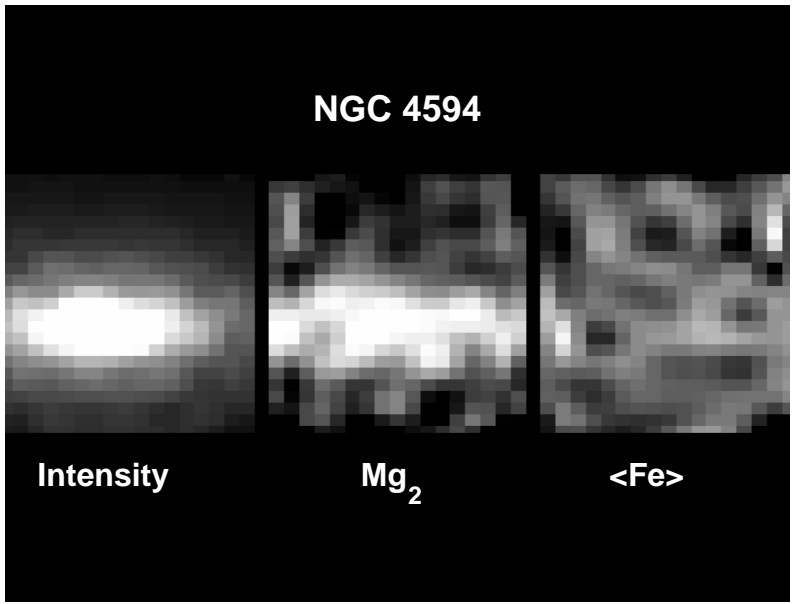}}
\caption{Absorption line maps of Mg$_2$ and $\langle$Fe$\rangle$ in the 
inner region of the Sombrero galaxy (field size is 8.2$''$ $\times$ 11$''$,
from Peletier et al. 1999). On the left a reconstructed continuum map is
shown. }
\end{center}
\end{figure}

The inner disk also shows an increased H$\gamma$ line strength, while
the CaT has a similar behaviour as $\langle$Fe$\rangle$ (Peletier 
et al. 1999). Although these data are still rather noisy for this purpose,
it is expected that much better data will become available soon from
the new, wide-field (33$''$ $\times$ 41$''$) integral field spectrograph
SAURON (PIs R. Bacon, P.T. de Zeeuw and R.L. Davies) on the WHT.

\section{What determines the Abundance Ratios in Galaxies?}

From the previous sections we can draw the following conclusions:

\begin{itemize}
\item Small galaxies (Mg$_{2,c}$ $<$ 0.22, $\sigma_c$ $<$ 150 km/s) have 
solar Mg/Fe ratios, large galaxies (Mg$_{2,c}$ $>$ 0.28, $\sigma_c$ $>$ 
225 km/s are overabundant in Mg.
\item There is no difference in the Mg/Fe ratio between ellipticals and
bulges of the same velocity dispersion (or Mg$_{2,c}$ value).
\item Within a galaxy Mg/Fe appears to remain constant.
\end{itemize}

It is thought that the reason the Mg/Fe ratio from galaxy to galaxy 
varies,  is that 
the ratio of the number of supernovae Type II vs. Type Ia can vary.
SNe Type II occur in massive stars, and produce large amounts of light
elements. SNe Type Ia come from binary accretion onto white dwarfs, 
and produce relatively much more Fe-peak elements (see e.g. Pagel 1998).
Since the lifetime of the progenitors of SNe Type II is very short,
there is a period of a few times 10$^8$ years from initial star 
formation in which enrichment through SNe Type II dominates (Worthey 1998).
The most popular scenarios to vary this ratio of the two SNe types as
a function of galaxy size (or velocity dispersion) are
\begin{itemize}
\item the formation time-scale, that should be shorter in large galaxies
\item a variation in the IMF, such that large galaxies have a relatively
larger fraction of massive stars.
\end{itemize}
Although it is difficult to reject the first scenario, there are various
reasons why I would prefer the second. If the Mg/Fe ratio would be 
determined by the formation time-scale, then bulges and ellipticals 
would have had to form the majority of their stars on the same time-scale.
Clearly this would predict that disks, in which the star formation is slow,
would have lower Mg/Fe ratios. Although no good measurements of disks are
available at present, the measurements by Fisher et al. (1996), which
show no difference in Mg/Fe between major and minor axis in S0 galaxies,
do not favour this scenario. Central disks (e.g. in the Sombrero)
have high, rather than low Mg/Fe ratios.
Secondly, it would be very difficult to
form the brightest galaxies (with high Mg/Fe ratios) in an hierarchical
way, on time-scales of Gyrs, since SNe Type Ia would lower the Mg/Fe of the 
gas, and make it very difficult to reach large Mg/Fe ratios.
For example, the central disk of the Sombrero should have formed very
fast from gas that was not very metal rich before the last merger event. 

The second option seems more favourable, although it also has its 
difficulties. If Mg/Fe would be a function just of one variable,
e.g., velocity dispersion (or escape velocity, as suggested by 
Franx \& Illingworth 1990), then this scenario would not be able 
to explain the line strength gradients in galaxies, since it seems 
that for the same Mg abundance Mg/Fe in the outer parts of bright galaxies 
is generally lower than Mg/Fe in the centres of faint galaxies (gradients
are steeper than the line connecting nuclei). To also be able
to explain the gradients we will modify this scenario: the IMF, which 
mainly determines the enrichment of the elements, has to be  
dependent only on the mass of the galaxy as a whole.
How realistic is this scenario? Although Elmegreen (these proceedings)
claims that the IMF is generally universal, there are indications in some 
starburst galaxies that the IMF there is skewed towards high mass stars 
(Kennicutt 1998, p. 71). At present we can't confirm, nor rule
out this second scenario. It is however capable of explaining the current
observations rather easily, much better than the first scenario.

The hot gas fraction, both in clusters of galaxies and in individual
objects, might further constrain chemical enrichment models of galaxies. 
The measurements in the ISM of elliptical galaxies however indicate
very low Fe-abundances (0.1 -- 0.4 times solar, Arimoto et al. 1997),
which seems inconsistent with the measurements from stellar spectra. 
For that reason we have to wait until the abundances from X-ray emssion
lines are better understood (see also Barnes, these proceedings).

\section{The Need for Better Stellar Models}

%\begin{itemize}
%\item non-solar abundance ratios
%\item fitting functions taking into account individual abundances
%\item opacities etc.
%\end{itemize}

Since abundance ratios in galaxies can only be obtained through detailed
comparisons with stellar models, it is crucial that the models are up to
date. At present, there are several stellar population models that 
predict integrated line strength indices on the Lick system. The models
do not differ too much from each other (Worthey 1994, Bruzual \& 
Charlot (see Leitherer et al. 1996), Vazdekis et al. 1996, Tantalo 
et al. 1996, Borges et al. 1995). Most of them use the stellar tracks
of the Padova group (Bressan, Chiosi \& Fagotto 1994). None of them however takes into
account the fact that if the abundance ratios in the stars are non-solar,
the stellar parameters, like effective temperature and gravity, might be
different. There have been some papers studying how stellar isochrones
change as a function of the Mg/Fe ratio for solar or larger metallicities
(Weiss, Peletier \& Matteucci 1995, Barbuy 1994). The conclusion of
Weiss et al. is that the isochrones basically do {\sl not} change, 
if the total metallicity (i.e. the mass fraction of elements heavier than 
He) remains constant. This is consistent with Barbuy (1994). 
If this result is reliable, isochrones with non-solar abundance ratios
are not necessary, as long as the abundance ratios are taken into
account in the fitting functions, which are used to calculate the line indices.
This result is in agreement with Salaris, Chieffi \& Straniero 
(1993) for lower
metallicities. However, in a new paper Salaris \& Weiss (1998), using
new opacities, find that for Z=0.01 (their largest metallicity)
$\alpha$-enhancement, even while keeping the total metallicity constant,
does change properties like the Main Sequence Turnoff and the RGB colour
significantly. It is important to find out why this result is 
in contradiction with previous work.

The results of Idiart et al. (1997), who showed that very different 
Ca-abundances are obtained for models of integrated stellar populations 
if the Ca-abundance of individual stars
is taking into account in the fitting functions, indicate that accurate
fitting functions are crucial. It is not expected that the models for 
$\langle$Fe$\rangle$ and Mg$_2$ will change much, since in Weiss et al. (1995)
it is shown that if one takes solar neighbourhood stars to calculate
the fitting functions for Mg$_2$ and $\langle$Fe$\rangle$, the integrated
indices are very similar to the ones that one obtains when determining the
fitting functions from Galactic Bulge stars. It is clear however, that
the next generation of stellar population model will have to include
fitting functions determined from as many stars as possible, using 
abundance ratios calculated  for each star individually. Using
8m telescopes, it should be possible to obtain spectra of stars covering 
the parameter space of temperature, gravity, metallicity and some abundance 
ratios, that would be required to do this.

\section{Summary}

I discuss the evidence for abundance ratios in galaxies, supplementing
the recent review by Worthey (1998). My main conclusions are:

\begin{itemize}
\item The scatter for early-type galaxies in the relations between 
$\langle$Fe$\rangle$ and H$\beta$ vs. velocity dispersion is small, 
comparable to the scatter in the Mg$_2$ and colour vs. $\sigma$ relation.
\item Small galaxies (Mg$_{2,c}$ $<$ 0.22, $\sigma_c$ $<$ 150 km/s) have 
solar Mg/Fe ratios, large galaxies (Mg$_{2,c}$ $>$ 0.28, $\sigma_c$ $>$ 
225 km/s are overabundant in Mg. There is no difference in the Mg/Fe ratio 
between ellipticals and bulges of the same velocity dispersion 
(or Mg$_{2,c}$ value). Within a galaxy Mg/Fe appears to remain constant.
\item We know very little known about the behaviour of other elements
(see Worthey 1998), although [Ca/Fe] in giant ellipticals appears to be solar, 
contrary to what one would expect from an $\alpha$-element. 
\item Improved stellar population models, using stellar evolutionary 
tracks with non-solar abundance ratios and fitting functions using 
abundance ratios determined for each standard star separately, would be
very welcome to calculate accurate abundance ratios.
\end{itemize}

\acknowledgments

I like to thank Claire Halliday, Alejandro Terlevich and Harald Kuntschner for 
communicating results in advance of publication, and John Beckman for organising
an interesting meeting.

\end{document}